\documentclass[12pt,preprint]{aastex}

\begin{document}

\title{An Observational Pursuit for Population III Stars in
       a Ly$\alpha$ Emitter at $z$=6.33
       through He{\sc ii} Emission\altaffilmark{1}}

\author{
          T. Nagao            \altaffilmark{2, 3},
          K. Motohara         \altaffilmark{4}, 
          R. Maiolino         \altaffilmark{2},
          A. Marconi          \altaffilmark{2}, 
          Y. Taniguchi        \altaffilmark{5}, \\
          K. Aoki             \altaffilmark{6}, 
          M. Ajiki            \altaffilmark{5}, and
          Y. Shioya           \altaffilmark{5}
}

\altaffiltext{1}{Based on data collected at the
         Subaru Telescope, which is operated by
         the National Astronomical Observatory of Japan.}
\altaffiltext{2}{INAF -- Osservatorio Astrofisico di Arcetri,
         Largo Enrico Fermi 5, 50125 Firenze, Italy}
\altaffiltext{3}{National Astronomical Observatory of Japan,
         2-21-1 Osawa, Mitaka, Tokyo 181-8588, Japan}
\altaffiltext{4}{Institute of Astronomy, Graduate School of Science,
         University of Tokyo, 2-21-1 Osawa, Mitaka, Tokyo 181-0015, Japan}
\altaffiltext{5}{Astronomical Institute, Graduate School of Science,
         Tohoku University, Aramaki, Aoba, Sendai 980-8578, Japan}
\altaffiltext{6}{Subaru Telescope, National Astronomical Observatory 
         of Japan, 650 N. A'ohoku Place, Hilo, HI 96720}

\begin{abstract}
We present a very deep near-infrared spectroscopic 
observation of a strong Ly$\alpha$ emitter at $z=6.33$, 
SDF J132440.6+273607, which we used to search for 
He{\sc ii}$\lambda$1640.  This emission line is expected 
if the target hosts a significant number of population III 
stars.  Even after 42 ksec of integration with the Subaru/OHS 
spectrograph, no emission-line features are detected in the 
$JH$ band, which confirms that SDF J132440.6+273607 is neither
an active galactic nucleus nor a low-$z$ emission-line object. 
We obtained a 2$\sigma$ upper-limit of $9.06 \times 10^{-18}$ 
ergs s$^{-1}$ cm$^{-2}$ on the He{\sc ii}$\lambda$1640 
emission line flux, which corresponds to a luminosity of 
$4.11 \times 10^{42}$ ergs s$^{-1}$. This upper-limit on the 
He{\sc ii}$\lambda$1640 luminosity implies that the upper 
limit on population III star-formation rate is in the range 
$4.9 - 41.2 M_\odot {\rm yr^{-1}}$ if population III stars 
suffer no mass loss, and in the range 
$1.8 - 13.2 M_\odot {\rm yr^{-1}}$ if strong mass loss is 
present. The non-detection of He{\sc ii} in SDF 
J132440.6+273607 at $z=6.33$ may thus disfavor weak feedback 
models for population III stars.
\end{abstract}

\keywords{
early universe ---
galaxies: evolution ---
galaxies: formation ---
galaxies: individual (SDF J132440.6+273607) ---
galaxies: starburst ---
stars: early-type}

\section{INTRODUCTION}

Population III stars, i.e., the first generation of stars, 
have been intensively investigated from the theoretical point 
of view. They are expected to have unique and distinct physical 
properties and to exhibit very different evolutionary processes 
compared with metal-enriched stars (e.g., Ezer \& Cameron 1971; 
Fujimoto et al. 1990; Bromm et al. 1999; Nakamura \& Umemura 2001; 
Abel et al. 2002; Omukai \& Palla 2003). Recent progress in the 
observational studies on the cosmic reionization have greatly 
increased the interests in the population III stars, as a possible 
important ionization source (e.g., Haiman \& Loeb 1997; 
Ciardi et al. 2003). Population III stars have also been 
investigated within the context of the chemical evolution of 
galaxies and quasars in their very early stages (e.g., 
Venkatesan et al. 2004; Matteucci \& Pipino 2005). Given these 
theoretical interests, the detection and the observational 
investigation of population III stars will be one of the main 
concerns of observational astronomy in the next decade. 

It should be noted that galaxies that contain a significant 
fraction of population III stars may have been already observed 
in the high-$z$ universe. Although the WMAP data suggest that 
the cosmic reionization occurred at $z \sim 17$ (Bennett et al. 
2003), this does not rule out the possibility of the existence 
of population III stars in the redshift range currently 
accessible (i.e., $z \lesssim 7$). Indeed Scannapieco et al. 
(2003) calculated the evolution of population III stellar 
systems and showed that these objects may exist even at 
$z \lesssim 7$, depending on the fundamental parameters of 
population III stars such as the initial-mass function and 
outflow efficiency. Therefore it is of interest to 
search for observational signatures from population III stars 
in the currently known very high-$z$ galaxies. 

Recently it has been recognized that the He{\sc ii}$\lambda$1640
emission is a good tracer for population III stars (e.g., 
Tumlinson \& Shull 2000; Oh et al. 2001; Schaerer 2002). The 
population III stellar clusters are expected to have very high 
effective temperature up to $\sim 10^5$K (e.g., Ezer \& Cameron 
1971; Bromm et al. 2001) as a result of both their top-heavy 
mass function and of small opacity of their atmosphere, due to 
the lack of heavy elements. Therefore population III stars emit 
a large fraction of photons at $h\nu > 54.4$eV, resulting in
large regions of doubly ionized helium. Moreover, since the 
ionized gas around population III stars does not contain heavy 
elements which are efficient coolants, He{\sc ii} is one of the 
main coolants and therefore the emissivity of He{\sc ii} lines 
is strongly enhanced. Among the He{\sc ii} lines, 
He{\sc ii}$\lambda$1640 is very important because this 
transition does not suffer from resonant absorption effects. 
Schaerer (2003) showed that the rest-frame equivalent width
of He{\sc ii}$\lambda$1640 can reach as much as a few tens 
${\rm \AA}$ if a galaxy contains a large number of population III
stars, which suggests that the detection of He{\sc ii}$\lambda$1640 
from population III stars at a high redshift may be feasible with
current observational facilities (see also Tumlinson et al. 2003).
Dawson et al. (2004) investigated the spectra of 17 
Ly$\alpha$ emitters (LAEs) at $z \approx 4.5$ and found no 
He{\sc ii}$\lambda$1640
signal, implying that population III stars should form at
$z > 4.5$.

Motivated by these expectations, we performed a 
deep near-infrared spectroscopic observation of a recently 
discovered LAE in the Subaru Deep Field 
(SDF; Kashikawa et al. 2004), SDF J132440.6+273607 at $z=6.330$
(Nagao et al. 2004) to search for the signature of population 
III stars through He{\sc ii}$\lambda$1640.
In this Letter, we present the result of our observation and
give some observational constraints on the presence of 
population III stars. We adopt a cosmology with 
($\Omega_{\rm tot}$, $\Omega_{\rm M}$, $\Omega_{\Lambda}$)
=(1.0, 0.3, 0.7) and $H_0$ = 70 km s$^{-1}$ Mpc$^{-1}$ 
throughout this Letter. 
We use the AB photometric system for the magnitude notation.

\section{OBSERVATIONS}

To search for population III stars through 
He{\sc ii}$\lambda$1640, we carried out a deep near-infrared 
spectroscopic observation of a LAE at $z=6.33$, 
SDF J132440.6+273607 (Nagao et al. 2004), 
by using the OH-airglow suppressor (OHS; 
Iwamuro et al. 2001) with the Cooled Infrared Camera and 
Spectrograph for OHS (CISCO; Motohara et al. 2002) installed 
on the Nasmyth focus of the 8.2m Subaru Telescope (Iye et al. 
2004), on 22--24 April 2005 (UT). This object was selected 
because it shows a relatively large Ly$\alpha$ luminosity 
(1.8 $\times 10^{43}$ ergs s$^{-1}$), high equivalent width 
[$EW_0 ({\rm Ly}\alpha) = 130 \pm 30 {\rm \AA}$], and because it is 
close to a relatively bright object ($i^\prime = 20.8$ and 
9$\farcs$0 east from the target) that helps a secure 
acquisition of the target onto the slit. This object has 
also the advantage that the redshifted He{\sc ii}$\lambda$1640 
line does not fall on the wavelength of a strong OH-airglow 
line, which will be highly suppressed by OHS. The finding 
chart is shown in Figure 1. Thanks to the OH-airglow 
suppression technique, OHS can achieve a very high sensitivity 
level for $JH$-band spectroscopy. The spatial and spectral 
samplings are 0$\farcs$105 and $\sim$8.5${\rm \AA}$ per pixel 
on the HAWAII 1K$\times$1K array. The $JH$-band spectra were 
acquired using a 0$\farcs$95 slit with a single exposure time 
of 1000 sec, which results into a wavelength resolution of 
$R \sim 200$ for the wavelength coverage of 
$1.108\mu {\rm m} < \lambda_{\rm obs} < 1.353\mu {\rm m}$ and
$1.477\mu {\rm m} < \lambda_{\rm obs} < 1.804\mu {\rm m}$.
During the observation, the telescope was nodded with a 
dithering of 4$\farcs$5 or 5$\farcs$5 for background 
subtraction. Although the total integration time was 54,000 
sec, we removed some low-quality frames (e.g., bad seeing) 
and thus used only 42,000 sec data in the following analysis 
and discussion. The adopted position angle was 93$\fdg$5, so 
that the target and the reference bright object were observed 
simultaneously (note that the effective slit length is 
$\sim 25^{\prime \prime}$). 
Though the position angle was not parallactic during the
observation, the effect of the atmospheric dispersion is
negligible in the observed wavelength range.
We also obtained the spectra of 
HIP 66935, HIP 68767, HIP 71468, HIP 75911 and HIP 83063 for 
the correction of telluric and instrumental transmission 
curves. The data were reduced following the standard procedure 
of dark subtraction, flat fielding, sky subtraction, bad-pixel
correction, residual sky subtraction, and correction for the
atmospheric and instrumental transmission curves.
Spectrophotometry was carried out using the spectra of
FS 138 taken after the observation.

\section{RESULTS}

The final two-dimensional spectrum is shown in Figure 2. 
Three very strong continuum features (one is positive and 
two are negative) are seen, which are the spectra of the
reference star observed simultaneously with the scientific 
target. Although the redshifted He{\sc ii}$\lambda$1640 is 
expected at $\lambda_{\rm obs} \approx 1.202 \mu$m, no 
emission-line feature is seen at the expected place on the 
two-dimensional spectrum as shown in Figure 2. The aperture 
size for spectral extraction is 9 pixels (0$\farcs$95). In 
Figure 3, the one-dimensional spectrum around the expected 
wavelength of the He{\sc ii}$\lambda$1640 emission is shown,
adopting a 5 pixel binning. The spectral extraction of
the target is performed by tracing the reference star.
If we assume that the He{\sc ii} 
line is unresolved, or equivalently, that the He{\sc ii} 
emission-line width is less than $\sim$2000 km s$^{-1}$, 
the estimated 2$\sigma$ upper limit on the 
He{\sc ii}$\lambda$1640 flux from SDF J132440.6+273607 is 
$9.06 \times 10^{-18}$ ergs s$^{-1}$ cm$^{-2}$, which 
corresponds to a luminosity of $4.11 \times 10^{42}$ ergs 
s$^{-1}$. Since the Ly$\alpha$ flux of the galaxy is 
$4.0 \times 10^{-17}$ ergs s$^{-1}$ cm$^{-2}$ (Nagao et al. 
2004), the 2$\sigma$ upper-limit on the 
He{\sc ii}$\lambda$1640/Ly$\alpha$ ratio is 0.23. This upper 
limit would become significantly smaller if taking into 
account the absorption by intergalactic matter on the 
Ly$\alpha$ flux, as discussed in \S4. 

In the discovery paper of SDF J132440.6+273607 (Nagao et al. 
2004), the possibility that this object is a [O{\sc ii}] 
emitter at $z \sim 1.39$ was not rejected completely 
because the spectral resolution was too low to resolve the 
[O{\sc ii}] doublet. However, a [O{\sc ii}] emitter at 
$z \sim 1.39$ should exhibit H$\beta$ emission at 
$\lambda_{\rm obs} \sim$ 1.159$\mu$m and a [O{\sc iii}] 
doublet at $\lambda_{\rm obs} \sim 1.185\mu$m and 
$\sim 1.197\mu$m. Since the $JH$ spectrum of SDF 
J132440.6+273607 shows neither the He{\sc ii}$\lambda$1640 
emission nor any other emission-line features,  the idea 
that this target is not a low-$z$ emission-line galaxy but
an object at $z=6.33$ is further supported. Also, the 
non-detection of C{\sc iv}$\lambda$1549 at 1.135$\mu$m 
suggests that SDF J132440.6+273607 is not an active 
galactic nucleus (AGN) at $z=6.33$ since 
C{\sc iv}$\lambda$1549 is one of the strongest UV emission 
lines seen in AGNs. The estimated 2$\sigma$ upper-limit 
flux on C{\sc iv}$\lambda$1549 is $1.8 \times 10^{-17}$ 
ergs s$^{-1}$ cm$^{-2}$ with the same extraction aperture
adopted for the measurement of He{\sc ii}$\lambda$1640.

Not only emission-line features, but any continuum emission
of SDF J132440.6+273607 is also undetected. The 2$\sigma$ 
upper-limit flux densities are $5.2 \times 10^{-20}$ ergs 
s$^{-1}$ cm$^{-2}$ ${\rm \AA}^{-1}$ at $1.20\mu$m and 
$3.4 \times 10^{-20}$ ergs s$^{-1}$ cm$^{-2}$ 
${\rm \AA}^{-1}$ at $1.64\mu$m. Here the same spectrum 
extraction aperture (0$\farcs$95) is adopted. These 
upper-limit values are consistent with the flux density 
measured by NB921 filter (centered at 9196${\rm \AA}$;
see Kashikawa et al. 2004; Taniguchi et al. 2005), 
$F_{1255} = 4.3 \times 10^{-20}$ ergs s$^{-1}$ cm$^{-2}$ 
${\rm \AA}^{-1}$ (Nagao et al. 2004), assuming
a flat UV continuum.

\section{DISCUSSION}

Due to the very hard spectral energy distribution (SED) of 
population III stars, a strong He{\sc ii}$\lambda$1640 
emission should be seen if population III stars exist in the 
observed system. We can estimate an upper limit on the number 
of population III stars by means of the obtained upper limit
on the He{\sc ii}$\lambda$1640 flux. Once a SED of a stellar 
cluster is given, the He{\sc ii}$\lambda$1640 luminosity is 
given by the following relation;
\begin{equation}
  L_{1640} = 
    c_{1640} \ (1 - f_{\rm esc}) \ Q({\rm He}^+) \ 
      \left( \frac{SFR_{\rm PopIII}}{\rm M_\odot \ yr^{-1}} \right) =
    f_{1640} 
      \left( \frac{SFR_{\rm PopIII}}{\rm M_\odot \ yr^{-1}} \right) ,
\end{equation}
where $c_{1640}$ is the He{\sc ii}$\lambda$1640 emission 
coefficient for Case B ($5.67 \times 10^{-12}$ ergs),
$f_{\rm esc}$ is the escape fraction of the ionizing photon,
$Q({\rm He}^+)$ is the He$^+$ ionizing photon flux, and 
$f_{1640}$ is defined as 
$f_{1640} = c_{1640} \ (1 - f_{\rm esc}) \ Q({\rm He}^+)$.
Assuming that the escape fraction is negligibly small, the 
expected He{\sc ii}$\lambda$1640 luminosity is determined 
only by the shape of the SED, which depends on the 
initial-mass function (IMF) and on the evolutionary processes 
of population III stars. Schaerer (2002) computed detailed 
evolutionary models for population III stars and presented 
their SEDs for given IMF and two extreme evolutionary cases, 
i.e., with and without strong mass loss. The evolution of the 
SED is strongly affected by mass loss because the effective 
temperature of population III stars increases if strong mass 
loss occurs (e.g., El Eid et al. 1983) while it is much cooler 
without (e.g., Castellani et al. 1983; Chieffi \& Tornambe 1984).
The $f_{1640}$ calculated by Schaerer (2002) and the derived 
upper limits on $SFR_{\rm PopIII}$ are summarized in Table 1.
The investigated IMFs are Salpeter's ones (i.e., 
$\propto M^{-2.35}$) with lower-mass cut off at 
$M_{\rm low} = 1 M_\odot$ or $50 M_\odot$ and with upper-mass 
cut off at $M_{\rm up} = 100 M_\odot$, $500 M_\odot$ or 
$1000 M_\odot$. Although a very large (and nearly meaningless) 
upper limit is derived from the model with 
$M_{\rm up} = 100 M_\odot$, this model seems far from 
describing the actual situation for population III stars. 
Recent theoretical studies suggest that the formation of 
population III stars is characterized by a much larger 
$M_{\rm up}$, at least a few hundreds $M_\odot$ (e.g., 
Bromm et al. 2002; Omukai \& Palla 2003). Therefore, the upper 
limit on $SFR_{\rm PopIII}$ is estimated to be in the range 
$4.9 - 41.2 M_\odot {\rm yr^{-1}}$ for the cases without mass 
loss and $1.8 - 13.2 M_\odot {\rm yr^{-1}}$ for the cases with 
strong mass loss, depending on $M_{\rm low}$ and $M_{\rm up}$.
On the other hand, the total SFR (including population III 
$and$ population I/II stars) estimated through the Ly$\alpha$
flux is $SFR_{\rm total} \approx 16 \ M_\odot$ yr$^{-1}$ 
(Nagao et al. 2004). However, this is likely a lower limit 
since the absorption due to intergalactic medium on the 
Ly$\alpha$ flux is not taken into account. For high-$z$ LAEs, 
more than half of the Ly$\alpha$ flux could be absorbed due to 
high opacity of intergalactic matter (e.g., Haiman 2002). 
Therefore, our non-detection of the He{\sc ii}$\lambda$1640 
emission suggests that the star-forming activity in
SDF J132440.6+273607 is not dominated by population III but
by normal population I/II stars, unless $M_{\rm low}$ is very 
low ($\sim 1 M_\odot$) and mass loss does not occur during 
the evolutionary process of population III stars. Some 
theoretical studies suggest that strong mass loss is expected 
due to electron-scattering opacity, because population III 
stars are expected to be radiating near to their Eddington 
limit (e.g., El Eid et al. 1983).

Although it is impossible to derive general conclusions on 
the nature of population III stars based only on this case 
study of SDF J132440.6+273607, we will now consider the 
possible implications of our result. Scannapieco et al. (2003) 
computed $f_{\rm III}(z)$, the probability that the star 
formation occurring in a high-$z$ galaxy is dominated by 
population III stars as a function of redshift and efficiency 
of metal-enriched outflow from high-$z$ starburst.
Since population III star formation cannot occur once the 
primeval gas is polluted by metal-enriched outflow from a 
neighboring system, $f_{\rm III}(z)$ strongly depends on the 
metal-ejection efficiency of the population III stars. However, 
the metal ejection from population III stars takes place only 
in a narrow mass range 
($140 M_\odot \lesssim M_\ast \lesssim 260 M_\odot$) by pair 
instability supernovae (SNe$_{\gamma\gamma}$; e.g., 
Heger \& Woosley 2002), since population III stars with 
$M_\ast \lesssim 140 M_\odot$ and $M_\ast \gtrsim 260 M_\odot$ 
collapse to black holes and thus no metal ejection occurs. 
Therefore $f_{\rm III}(z)$ also depends on the initial-mass 
function (IMF) of population III stars. Scannapieco et al. 
(2003) parameterized the feedback effects with the parameter 
$\mathcal{E}_{\it g}^{\rm III}$, ``energy input per gas mass'', 
defined as $\mathcal{E}_{\it g}^{\rm III} = 
\mathcal{E}_{\rm kin} {\it f}_\ast {\it f}_{\rm w} \mathcal{N}^{\rm III}$
where $\mathcal{E}_{\rm kin}$, ${\it f}_\ast$, ${\it f}_{\rm w}$ 
and $\mathcal{N}^{\rm III}$ are, respectively, the explosion 
energies of SNe$_{\gamma\gamma}$, the fraction of gas 
converted into stars, the fraction of $\mathcal{E}_{\rm kin}$ 
channeled into galaxy outflows, and the number of 
SNe$_{\gamma\gamma}$ per unit stellar mass. Scannapieco et al. 
(2003) reported that the population III star formation could 
continue to $z=6.33$ only in the case of weak feedback models 
(i.e., small $\mathcal{E}_{\it g}^{\rm III}$). Our non-detection 
of population III stars at $z=6.33$ would disfavor the models 
with $\mathcal{E}_{\it g}^{\rm III} \lesssim$ $10^{47.5}$ ergs 
$M_\odot^{-1}$, which corresponds to population III IMFs with a 
narrow Gaussian peak at $M_\ast \lesssim 100 
M_\odot$ or $M_\ast \gtrsim 500 M_\odot$.

To make the above discussion more general, He{\sc ii} emission 
should be investigated in many more galaxies at 
$6 \lesssim z \lesssim 7$ and, indeed, the most recent 
observational facilities have allowed us to find some galaxies 
in this redshift redshift range with spectroscopically confirmed 
redshifts (e.g., Hu et al. 2002; Kodaira et al. 2003; Kurk et al. 
2004; Rhoads et al. 2004; Nagao et al. 2004, 2005; Stern et al. 
2005; Taniguchi et al. 2005). Among them, galaxies with a large 
$EW_0$(Ly$\alpha$)  are good targets because galaxies hosting a 
significant number of population III stars are expected to 
present very large equivalent widths 
[$EW_0({\rm Ly}\alpha) \gtrsim 1000{\rm \AA}$; e.g., Schaerer 
2002; Scannapieco et al. 2003]. However, it may be more 
promising to search for population III stars at much higher 
redshifts because the probability that galaxies are dominated 
by population III stars increases dramatically as a function 
of $z$, as demonstrated by Scannapieco et al. (2003).
Recently, the near-infrared excess of the cosmic background 
radiation (Wright \& Reese 2000; Cambr\'{e}sy et al. 2001; 
Matsumoto et al. 2005) has been interpreted to be the 
integrated emission from population III stars in their final 
phase of formation at $z \sim 9$ (e.g., Santos et al. 2002;
Salvaterra \& Ferrara 
2003). This implies that the most promising way to 
observationally detect and investigate population III stars
is surveying strong Ly$\alpha$ emitters at $z\sim9$ and examine 
their He{\sc ii} emission. Although a recent Ly$\alpha$ emitter 
survey failed to find such objects (Willis \& Courbin 2005), 
forthcoming infrared cameras with a large FOV such as MOIRCS 
(Tokoku et al. 2003) will be powerful tools to detect  
Ly$\alpha$ emitters in the very high-$z$ universe. 

\vspace{0.5cm}

We thank the Subaru Telescope staffs for their invaluable assistance.
We also thank F. Iwamuro for the useful discussion on the OHS data 
reduction, and K. Omukai and R. Schneider for their 
valuable comments. TN and MA are JSPS fellows.

\clearpage
%-------------------------------------------------------------------------

\clearpage

\begin{deluxetable}{cccc}
\tablenum{1}
\tablecaption{Upper Limits on Population III Star-Formation Rate}
\tablewidth{0pt}
\tablehead{
 \multicolumn{2}{c}{Model\tablenotemark{a}} &
 \colhead{$f_{1640}$\tablenotemark{b}} &
 \colhead{$SFR_{\rm PopIII}$\tablenotemark{c}} \\
 \colhead{IMF} &
 \colhead{Mass Loss} &
 \colhead{(ergs s$^{-1}$)} &
 \colhead{(M$_\odot$ yr$^{-1}$)}
}
\startdata
$1 \lesssim M_\ast/M_\odot \lesssim 100$ & no
  & $1.91 \times 10^{40}$ &  $<$ 215.2 \\
$1 \lesssim M_\ast/M_\odot \lesssim 500$ & no
  & $9.98 \times 10^{40}$ &  $<$ 41.2 \\
$50 \lesssim M_\ast/M_\odot \lesssim 500$ & no
  & $8.38 \times 10^{41}$ &  $<$ 4.9 \\
$1 \lesssim M_\ast/M_\odot \lesssim 500$ & yes
  & $3.12 \times 10^{41}$ &  $<$ 13.2 \\
$50 \lesssim M_\ast/M_\odot \lesssim 1000$ & yes
  & $2.33 \times 10^{42}$ &  $<$ 1.8 \\
\enddata
\tablenotetext{a}{Parameters adopted by Schaerer (2002).}
\tablenotetext{b}{From Schaerer (2002).}
\tablenotetext{c}{2$\sigma$ upper-limit on SFR of population III stars
   derived by our upper limit on $F$(He{\sc ii}$\lambda$1640).}
\end{deluxetable}

\clearpage

\begin{figure*}
\epsscale{0.80}
\plotone{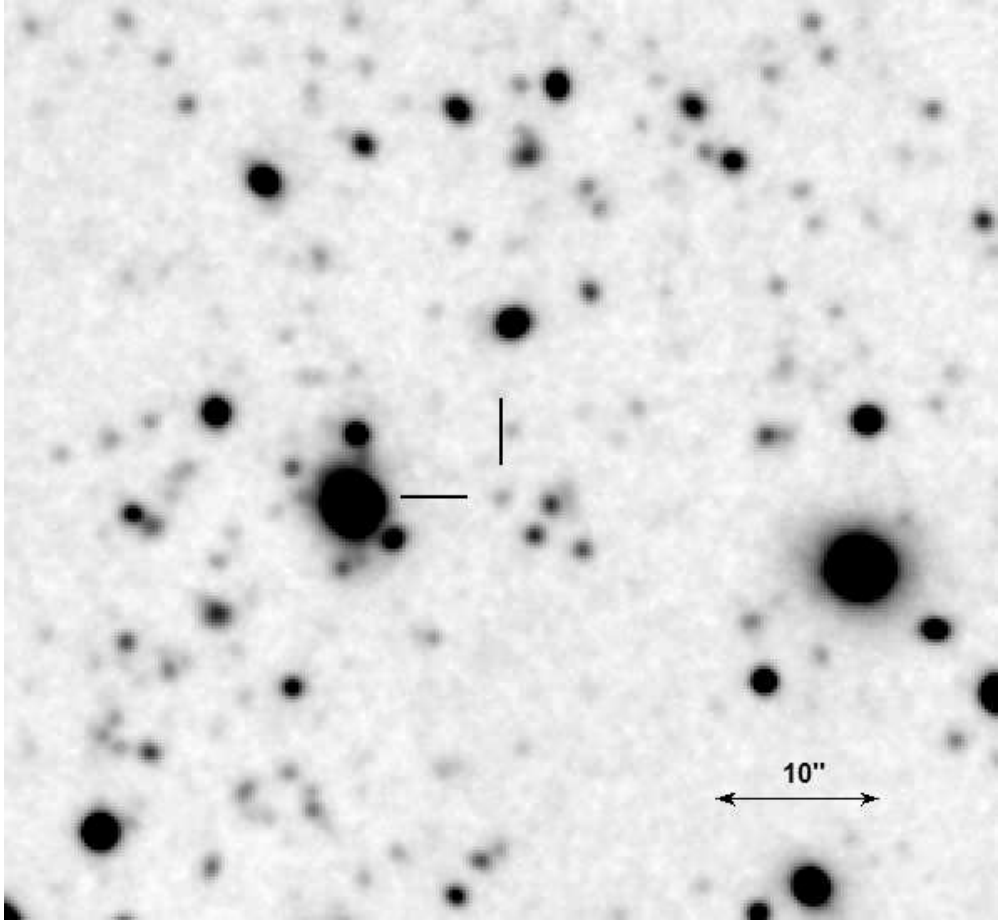}
\caption{
Finding chart of our target object, SDF J132440.6+273607
(center), taken from the SDF $z^\prime$-band image.
North is up and east is left. The horizontal arrow denotes 
an angular scale of $10^{\prime \prime}$, and the total 
angular size of the displayed area is 
$60^{\prime \prime} \times 60^{\prime \prime}.$
\label{fig1}}
\end{figure*}

\clearpage

\begin{figure*}
\vspace{-95mm}
\epsscale{1.05}
\plotone{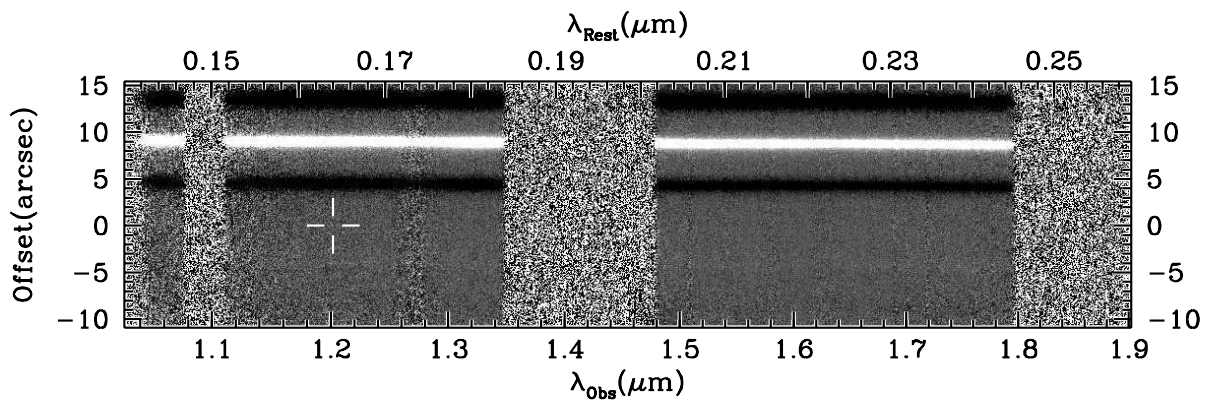}
\caption{
Two-dimensional $JS$-band spectrum of SDF J132440.6+273607.
Rest frame and observed wavelengths are given in the upper 
and lower sides of the panel, respectively. The spatial scale
along the slit is given at the both sides of the panel.
East is up and west is down. The expected wavelength of the 
redshifted He{\sc ii}$\lambda$1640 emission of 
SDF J132440.6+273607 ($\lambda_{\rm obs} = 1.202\mu$m) is 
marked by the white vertical lines, and the expected position of
the target is denoted by the white horizontal lines.
Vertical gaps seen at $\lambda_{\rm obs} \sim 1.1 \mu$m,
$\sim 1.4 \mu$m and $\gtrsim 1.8 \mu$m are the gaps of
the OHS sensitivity.
\label{fig2}}
\end{figure*}

\clearpage

\begin{figure*}
\epsscale{1.00}
\plotone{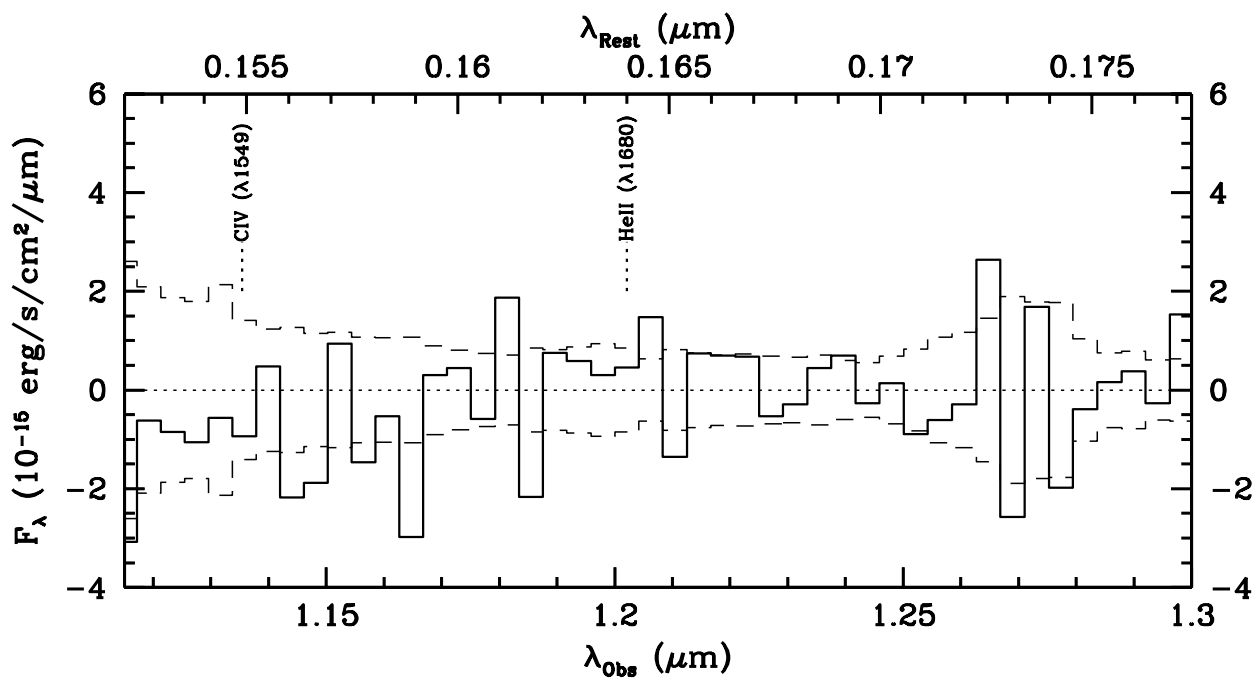}
\caption{
One-dimensional spectrum of SDF J132440.6+273607 with a
5 pixel binning (i.e., $\sim$1000 km s$^{-1}$).
The vertical arrows denote the expected wavelength of the
C{\sc iv}$\lambda$1549 emission ($\lambda_{\rm obs} = 1.135 \mu$m;
expected in the case of AGNs), and that of the 
He{\sc ii}$\lambda$1640 emission ($\lambda_{\rm obs} = 1.202 \mu$m;
expected for population III stars at $z=6.33$). 
The dotted line denotes the zero level of the spectrum, and
the dashed lines denote the 1$\sigma$ error level.
\label{fig3}}
\end{figure*}

\end{document}